\newif\iffigs
\figstrue
\documentclass[12pt]{article}
\usepackage{colordvi}
\usepackage{latexsym,amssymb}
\iffigs
    \usepackage{graphicx}
    \input{epsf}
\else
\fi
\newcommand{\ft}[2]{{\textstyle\frac{#1}{#2}}}

\newsavebox{\uuunit}
\sbox{\uuunit}
                             {\setlength{\unitlength}{0.825em}
                                      \begin{picture}(0.6,0.7)
                                                                  \thinlines
                                                                  \put(0,0){\line(1,0){0.5}}
                                                                  \put(0.15,0){\line(0,1){0.7}}
                                                                  \put(0.35,0){\line(0,1){0.8}}
                                                                 \multiput(0.3,0.8)(-0.04,-0.02){10}{\rule{0.5pt}{0.5pt}}
                                      \end {picture}}

\makeatletter \@addtoreset{equation}{section} \makeatother

\newtheorem{definizione}{Principle}[section]
\def\bfone{\relax{\rm 1\kern-.35em 1}}

\newif\ifpdf
\ifx\pdfoutput\undefined
   \pdffalse
   \usepackage{cite}
 \else
   \pdfoutput=1
   \pdftrue
  \usepackage[pdftex]{hyperref}
\fi
\begin{document}
\begin{titlepage}
\begin{flushright}
DISTA-2008 
\end{flushright}
\vskip 1.5cm
\begin{center}
{\Large \bf Free Differential Algebras, Rheonomy\\
\vskip 0.1cm
and Pure Spinors
} \\
 \vfill
{\bf Pietro Fr{\'e}$^1$ and Pietro Antonio Grassi$^2$}
\vfill {
$^1$ Dipartimento di Fisica Teorica, Universit{\`a} di Torino, \\
$\&$ INFN -
Sezione di Torino\\
via P. Giuria 1, I-10125 Torino, Italy\\
$^2$ DISTA, Universit\`a del Piemonte Orientale, \\
{ Via Bellini 25/G,  Alessandria, 15100, Italy
$\&$ INFN - Sezione di
Torino
\vskip .3cm
}
}
\end{center}
\vfill
\begin{abstract}
We report on progresses on the derivation of pure spinor constraints,
BRST algebra and BRST invariant sigma models a la pure spinors from
the algebraic structure of the FDA underlying supergravity.
\end{abstract}
\vfill
\vspace{1.5cm}
\vspace{2mm} \vfill \hrule width 3.cm {\footnotesize
$\null$\\
Talk given by Pietro Fr\'e at the Workshop Supersymmetry and Quantum
Symmetry 2007 held at the Joint Institute for Nuclear Research in
Dubna (Russian Federation), July 2007
}
\end{titlepage}
\section{Introduction}
A fundamental problem ubiquitous in current research on string theory
is that of calculating string amplitudes in the presence of background
Ramond Ramond fields. Neither the  Neveu Schwarz nor the  Green Schwarz formulation of
the string sigma model is  apt to such a task because these formulations treat 
the Ramond-Ramond background fields either non-polynomially (by means of spin fields) or 
non-covariantly (using light-cone gauge). An additional problem in the case
of the Green Schwarz formulation is located in the BRST quantization
of its distinctive local symmetry namely $\kappa$-supersymmetry. The
non-conventional nature of this symmetry which classically is nothing
else but the pull-back on the world-sheet of half of the
supersymmetry transformations of bulk supergravity, gives origin to
an infinite hierarchy of ghosts for ghosts.
\par
An apparently thorough solution of all these problems has been
provided by the Berkovits' pure spinor reformulation of the string
sigma model \cite{Berkovits:2000fe}. Berkovits' construction is an extended version of the
Green Schwarz formulation. In addition to the classical fields it
introduces a triplet of new fields, $\lambda^i, w_i, d_i$ which are
all target space $32$-components spinors. The last two members of the triplet $w_i$, $d_i$ are also
world-volume vectors as denoted by the index $i$. This triplet is the
classical one of BRST quantization, the three members being respectively
characterized by ghost number $1,-1,0$  and admitting
therefore the interpretation of ghost, antighost and Lagrange
multiplier. Indeed in Berkovits approach the starting point is
provided by the definition of a  BRST operator :
\begin{equation}
  Q^{BRST} \, = \, \int \, \lambda^i \, d_i  \,  d^2\sigma^i
\label{BRSTQ}
\end{equation}
against which the string sigma-model is required to be invariant. At
the same time the BRST charges are requested to be nilpotent.
Although only in a sense yet to be clarified, it is clear of which symmetry the
$\lambda$-fields, which are commuting, are supposed to be the ghosts:
this is bulk supersymmetry pull-backed onto the world sheet. Just as
$\kappa$-supersymmetry. The catch of the method is provided by the
constraints of the type:
\begin{equation}
  \overline{\lambda} \, \Gamma^a \, \lambda \, \approx \, 0
\label{Psconst1}
\end{equation}
which the superghosts $\lambda$ are requested to satisfy. In
Berkovits approach the  constraints of type (\ref{Psconst1}), named
pure spinor constraints are an a-priori input which constitutes part
of the definition of the BRST operator. At a second stage of
development, in Berkovits approach, one tries to determine the
constraints on target superspace geometry required for BRST
invariance of the sigma model action. For full consistency of the
approach it should happen that these latter constraints be those
describing target space supergravity. Although this can be achieved
through elaborate steps \cite{Berkovits:2001ue}, yet in this approach a clearcut correspondence between the
background bulk geometry and the pure spinor sigma model is not
available. This  makes it difficult to perform an
immediate direct construction of the pure spinor sigma model for any
chosen supergravity background. This becomes particularly evident
when one considers backgrounds with a reduced number of conserved
supersymmetries like $\mathrm{AdS_5} \times \mathrm{T^{1,1}}$ for the type IIB case,
 the compactification on $\mathrm{AdS_3} \times
\mathrm{CP^3}$ for the type IIA theory.
\par
It must also be noted that the pure spinor constraints
(\ref{Psconst1}) are of the general type indicated but, depending on
the chosen theory (type IIA, type IIB or M-theory) and on the chosen
background, have to be tuned in their explicit form.
\par
For this reasons it would be highly desirable to have a formulation
of the pure spinor sigma models in which the pure spinor constraints,
the BRST operator and the entire set up follow from background
supergravity just as it happens for the $\kappa$-supersymmetric
actions.
\par
Such a formulation is in progress. Work has been done in \cite{antonpietro,Fre:2007xy}
in the case of the M2-brane and new results are upcoming for the case
of type II superstrings \cite{pi_ant_ma_ri,pie_ant3}.
\par
The general idea is encoded in the following list of constructive
steps:
\begin{table}[!hbt]
\begin{center}
\iffigs
 \includegraphics[height=125mm]{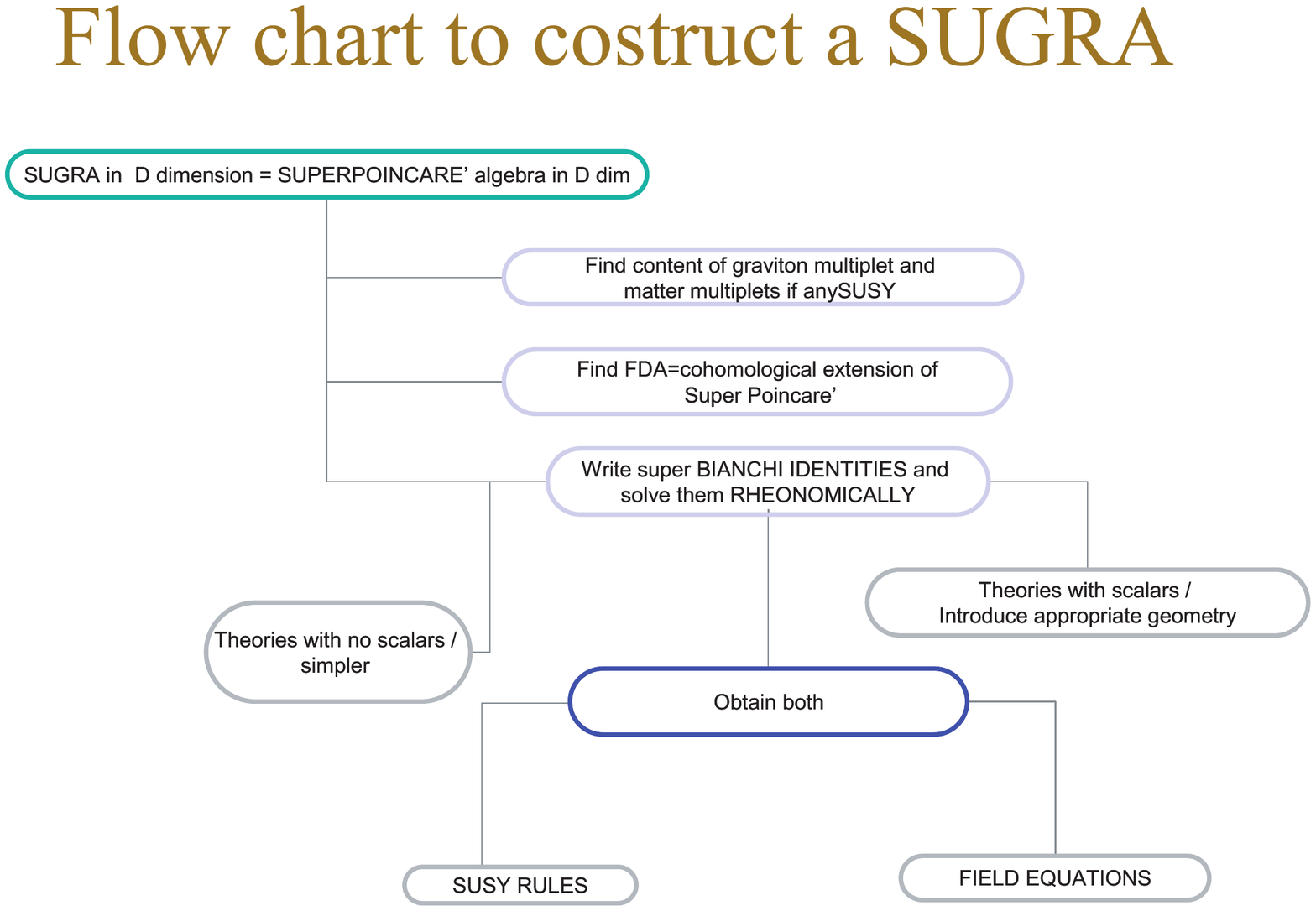}
\else
\end{center}
 \fi
\caption{\it This figure illustrates the flow chart for the construction of supergravity
theories. From the superalgebra to the field equations and supersymmetry transformation rules
everything is encoded in the structure constants of the superalgebra plus the principle of rheonomy}
\label{flowchart}
 \iffigs
 \hskip 1.5cm \unitlength=1.1mm
 \end{center}
  \fi
\end{table}
\begin{enumerate}
\item The algebraic structure underlying any higher dimensional supergravity theory
is a Free Differential Algebra (FDA). This latter is a categorical
extension of a (super) Lie algebra determined by the Chevalley
cohomology of this latter.
\item Given the FDA one considers its Bianchi identities and
constructs the unique rheonomic parametrization of the FDA
curvatures. Rheonomy is a universal principle of analiticity in
superspace (see fig.\ref{rheotable})which requires that the fermionic components of the FDA
curvatures should be linear functions of their bosonic ones. Rheonomy
encodes in one single principle the construction of both field equations
and supersymmetry transformation rules for any supergravity. Indeed
field equations follow as integrability conditions of the rheonomic
parametrization of curvatures. The flow chart for the construction of
classical supergravities and the principle of rheonomy are
respectively illustrated in table \ref{flowchart} and \ref{rheotable}
\item Consider then the FDA appropriate to the supergravity under
investigation and the rheonomic parametrization of its curvatures.
\item Perform the ghost-form extension of the classical FDA
according to the principle introduced by Anselmi and Fr\'e in \cite{Anselmi:1992tj}, namely:
\begin{definizione}\label{pippo}
$\null$\\
The correct BRST algebra is provided by replacing, in the rheonomic parametrization, of the classical
supergravity curvatures each
differential form with its extended ghost-form counterpart while keeping the curvature components
untouched.
Thus one obtains the rheonomic parametrization of the ghost--extended curvatures, whose formal definition
is identical with that of the classical curvatures upon the replacements:
\begin{equation}
  \begin{array}{ccc}
    d & \mapsto & d+\mathcal{S} \\
    \Omega^{[n]} & \mapsto & \sum_{p=0}^n \, \Omega^{[n-p,p]} \
  \end{array}
\label{orletto1}
\end{equation}
\end{definizione}
 In this
way one has the ordinary (unconstrained) BRST algebra of supergravity.
\item Set to zero all the bosonic ghosts. This defines a
constrained BRST algebra and for consistency a certain set of
\textbf{pure spinor} constraints. The correct constraints are the
  projection onto the world-sheet (brane world volume) of these
  constraints. The pure spinor constraints should not be chosen a
  priori as an input.\footnote{We point out that the relation between pure spinor formulation and 
  extended supersymmetry algebras \cite{fredauria} has been discovered in 
  \cite{Grassi:2001ug}. Recenlty, also N. Berkovits pointed out the relation between the rheonomic 
  parametrization and the pure spinor superspace \cite{Berkovits:2006ik}.
  }
  \item Verify that the pure spinor constraints can be solved in
  terms of as many independent degrees of freedom as it is required
  for a $c=0$ conformal theory in d=2 in the case of superstrings.
  \item Introduce the appropriate antighosts and Lagrange multiplier
  field and construct the BRST invariant quantum action.
\end{enumerate}
\begin{table}[!hbt]
\begin{center}
\iffigs
 \includegraphics[height=105mm]{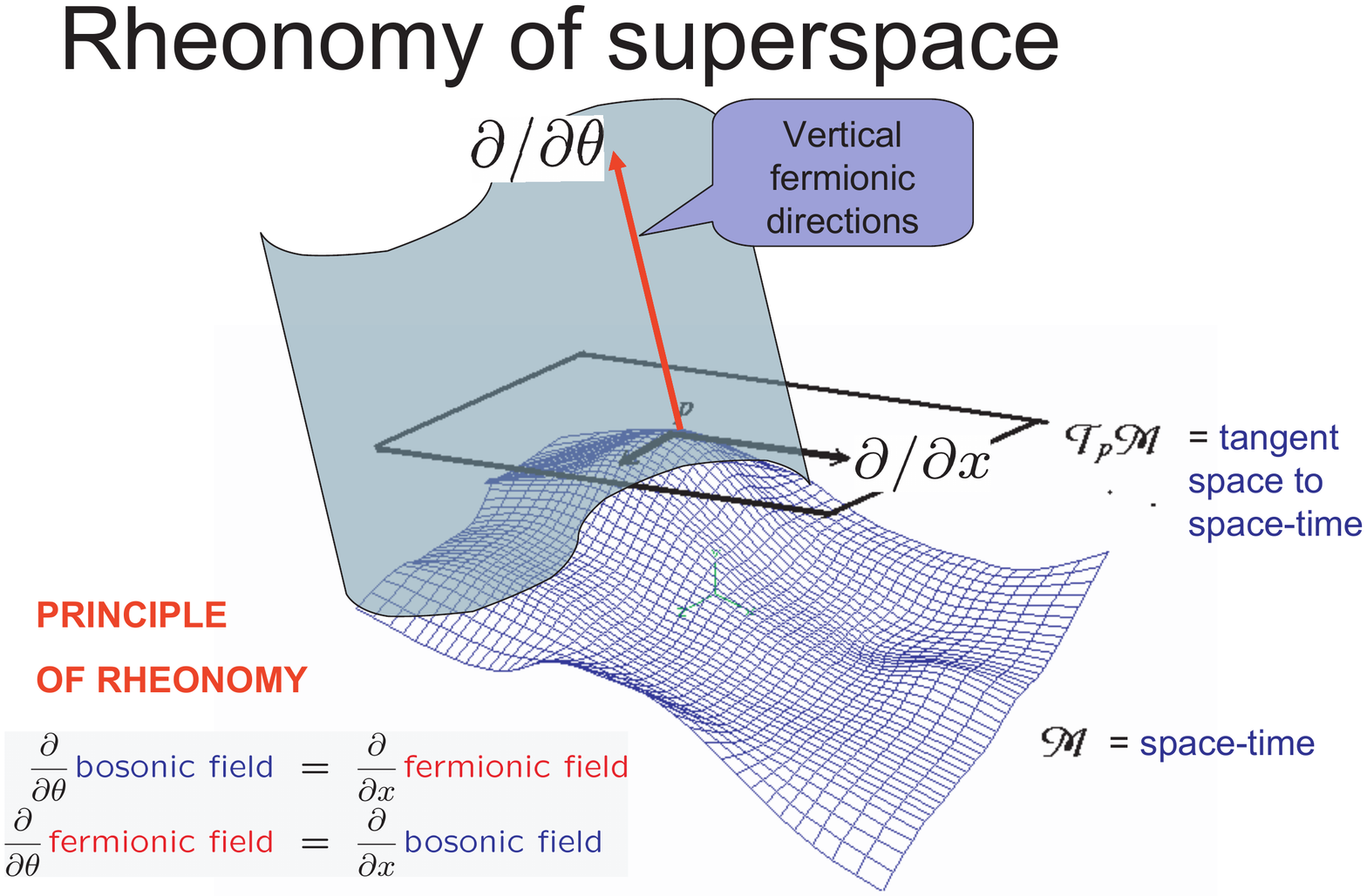}
\else
\end{center}
 \fi
\caption{\it The principle of rheonomy is an analogue of the Cauchy Riemann equations for analytic functions.
Just as the derivative of the imaginary part $v$ of $f(z)$ in the $x$ direction is related to the derivative of the
real part $u$ in the $y$ direction, in the same way the fermionic derivative of bosonic fields is expressed
as a combination of the bosonic derivatives of the fermionic ones. This is summarized by requiring that all external
components (fermionic) of the FDA curvature should be given as linear functions of  the inner components of the
 FDA curvatures. There is also an analogue of the differential equation satisfied by $u$ and $v$.
 This analogue are the field equations of supergravity which follow as integrability conditions of the rheonomic
 parametrization of FDA curvatures }
\label{rheotable}
 \iffigs
 \hskip 1.5cm \unitlength=1.1mm
 \end{center}
  \fi
\end{table}
In this talk we review the application of this scheme to the case of
M-theory (having in mind the M2-brane which was discussed in our common paper
\cite{antonpietro}). We use this example to
illustrate the flow chart of the construction. In particular we focus
on points 1-6 of the above list. Our main goal is to show how the
structure of the BRST algebra for all the fields with non negative
ghost number together with the pure spinor constraints are completely determined by the original superPoincar\'e
algebra through the following algorithmic steps which at each level yield the unique result displayed in table
\ref{downhill}:
\begin{table}[!hbt]
  \centering
  $$
  \begin{array}{c}
    \mbox{superPoincar\'e algebra} \\
    \Downarrow  \\
    \mbox{FDA} \\
    \Downarrow \\
    \mbox{Rheonomic solution of FDA Bianchis} \\
    \Downarrow \\
    \mbox{BRST ghost-extension} \\
    \Downarrow \\
    \mbox{Restriction to fermionic ghosts} \\
     \Downarrow \\
\mbox{Berkovits algebra and pure spinor constraints}\\
  \end{array}
  $$
  \caption{The deterministic path from the super Poincar\'e algebra to the constrained BRST algebra
   with pure spinors}\label{downhill}
\end{table}
The discussion of the antighost sector and of the BRST invariant
action is not treated here. It is the subject of the forthcoming
papers \cite{pi_ant_ma_ri,pie_ant3}
\section{General Structure of FDA.s and Sullivan's theorems}
\label{sullivano}
Free Differential Algebras (FDA) are a natural categorical extension of the
notion of Lie algebra and constitute the natural mathematical
environment for the description of the algebraic structure of higher
dimensional supergravity theory, hence also of string theory. The
reason is the ubiquitous presence in the spectrum of
string/supergravity theory of antisymmetric gauge fields ($p$--forms)
of rank greater than one.
\par
FDA.s were independently discovered in Mathematics by Sullivan \cite{sullivan}
and in Physics by the present author in collaboration with R. D'Auria
\cite{fredauria}. The original name given to this algebraic structure
by D'Auria and me was that of \textit{Cartan Integrable Systems}.
Later, recognizing the conceptual identity of our supersymmetric construction with the pure
bosonic constructions considered by Sullivan, we also turned to its naming FDA which has by now become
generally accepted.
\par
Let me recall
the definition of FDA.s and two structural theorems by Sullivan which
show how all possible FDA.s are, in a sense to be described,
\textit{cohomological extensions} of normal Lie algebras or
superalgebras.
\par
Another question which is of utmost relevance in all physical
applications is that of \textit{gauging} of FDA.s. Just in the same
way as physics gauges standard Lie algebras by means of Yang
Mills theory through the notion of gauge connections and curvatures one
expects to gauge FDA.s by introducing  their curvatures. A
surprising feature of the FDA setup which was noticed and explained
by me in a paper of 1985 \cite{comments} is that differently from Lie
algebras the algebraic structure of FDA already encompasses both the
notion of connection and the notion of curvature and there is a
well defined mathematical way of separating the two which relies on
the two structural theorems by Sullivan.
\par
\paragraph{Definition of FDA} The starting point for FDA.s is the
generalization of Maurer Cartan equations. A standard Lie algebra is defined by its
structure constants which can be alternatively introduced either
through the commutators of the generators:
\begin{equation}
  \left[ T_I \, , T_K\right] \, = \, \tau^{I}_{\phantom{I}JK} \, T_I
\label{commrel}
\end{equation}
or through the Maurer Cartan equations obeyed by the dual $1$--forms:
\begin{equation}
  d e^I = \ft 12 \, \tau^{I}_{\phantom{I}JK} \, e^J \wedge e^K
\label{MC1}
\end{equation}
The relation between the two descriptions is
provided by the duality relation:
\begin{equation}
  e^I(T_J) = \delta^I_J
\label{dualityrele}
\end{equation}
 Adopting
the Maurer Cartan viewpoint FDA.s can now be defined as follows.
Consider a formal set of exterior forms $\left\{ \theta^{A(p)}\right\}
$ labelled by the index $A$ and by the degree $p$ which may be
different for different values of $A$. Given this set we can write a
set of generalized \textit{Maurer Cartan equations} of the following type:
\begin{equation}
  d\theta^{A(p)} \, + \, \sum_{n=1}^{N} \,
  C^{A(p)}_{\phantom{{A(p)}}B_1(p_1)\dots B_n(p_n)} \,
  \theta^{B_1(p_1)} \, \wedge \, \dots \, \wedge \theta^{B_n(p_n)}\, = \,0
\label{detheta}
\end{equation}
where $C^{A(p)}_{\phantom{{A(p)}}B_1(p_1)\dots B_n(p_n)}$ are generalized structure
constants with the
  same symmetry as induced by permuting the $\theta$.s in the wedge
  product. They can be non--zero only if:
\begin{equation}
  p+1= \sum_{i=1}^n \, p_i
\label{p+1}
\end{equation}
Equations (\ref{detheta}) are self-consistent and define an FDA if
and only if $dd\theta^{A(p)}=0$ upon substitution of (\ref{detheta})
into its own derivative. This procedure yields the generalized Jacobi
identities of FDA.s.\footnote{For a review of FDA theory see
\cite{castdauriafre}}
\paragraph{Classification of FDA and the analogue of Levi theorem:  minimal versus contractible algebras} A
basic theorem of Lie algebra theory states that the most general Lie
algebra $\mathcal{A}$ is the semidirect product of a semisimple Lie
algebra $\mathcal{L}$ called the Levi subalgebra with
$\mbox{Rad}(\mathcal{A})$, namely with the radical of $\mathcal{A}$. By
definition this latter is the maximal solvable ideal of $\mathcal{A}$.
Sullivan \cite{sullivan} has provided an analogous structural theorem
for FDA.s. To this effect one needs the notions of \textit{minimal
FDA} and \textit{contractible FDA}. A minimal FDA is one for which:
\begin{equation}
  C^{A(p)}_{\phantom{A(p)}B(p+1)} \, = \, 0
\label{minimalconst}
\end{equation}
This excludes the case where a $(p+1)$--form appears in the
generalized Maurer Cartan equations as a contribution to the
derivative of a $p$--form. In a minimal algebra all non differential
terms are products of at least two elements of the algebra, so that
all forms appearing in the expansion of $d\theta^{A(p)}$ have at most
degree $p$, the degree $p+1$ being ruled out.
\par
On the other hand a \textit{contractible FDA} is one where the only
form appearing in the expansion of $d\theta^{A(p)}$ has degree $p+1$,
namely:
\begin{equation}
  d\theta^{A(p)} = \theta^{A(p+1)} \quad \Rightarrow \quad
  d\theta^{A(p+1)} = 0
\label{contraFDA}
\end{equation}
A contractible algebra has a trivial structure. The basis $\left\{
\theta^{A(p)}\right\}$ can be subdivided in two subsets $\left\{ \Lambda^{A(p)}\right\} $
and $\left\{ \Omega^{B(p+1)}\right\} $ where $A$ spans a subset of the values taken by
$B$, so that:
\begin{equation}
  d\Omega^{B(p+1)} = 0
\label{forallB}
\end{equation}
for all values of $B$ and
\begin{equation}
   d\Lambda^{A(p)}= \Omega^{A(p+1)}
\label{unguedA=B}
\end{equation}
Denoting by $\mathcal{M}^k$ the vector space generated by all forms
of degree $p\le k$ and $C^k$ the vector space of forms of degree $k$,
a minimal algebra is shortly defined by the property:
\begin{equation}
  d\mathcal{M}^k \, \subset \, \mathcal{M}^k \wedge \mathcal{M}^k
\label{lottominimo}
\end{equation}
while a contractible algebra is defined by the property
\begin{equation}
  d C^k \, \subset \, C^{k+1}
\label{lottocontrat}
\end{equation}
In analogy to Levi's theorem, the first theorem by Sullivan states
that: \textit{The most general FDA is the semidirect sum of a
contractible algebra with a minimal algebra}
\par
\paragraph{Sullivan's first theorem and the gauging of FDA.s}
Twenty years ago in \cite{comments} one of us  observed that the above
mathematical theorem has a deep physical meaning relative to the
gauging of algebras. Indeed I proposed the following identifications:
\begin{enumerate}
  \item The \textit{contractible generators} $\Omega^{A(p+1)}+\dots$ of any given FDA $\mathbb{A}$ are
  to be physically identified with the \textit{curvatures}
  \item The Maurer Cartan equations that begin with
  $d\Omega^{A(p+1)}$ are \textit{the Bianchi identities}.
  \item The algebra which is gauged is the \textit{minimal
  subalgebra} $\mathbb{M} \subset \mathbb{A} $.
  \item{The Maurer Cartan equations of the minimal subalgebra
  $\mathbb{M}$ are consistently obtained by those of $\mathbb{A}$ by setting
  all contractible generators to zero.}
\end{enumerate}
\par
\paragraph{Sullivan's second structural theorem and Chevalley cohomology}
The second structural theorem proved by Sullivan\footnote{For detailed explanations on this see
again, apart from the original article \cite{sullivan} the book \cite{castdauriafre}} deals with the
structure of minimal algebras and it is constructive. Indeed it
states that the most general minimal FDA $\mathbb{M}$ necessarily
contains an ordinary Lie subalgebra $\mathbb{G} \subset \mathbb{M}$
whose associated $1$--form generators we can call $e^I$, as in
equation (\ref{MC1}). Additional $p$--form generators $A^{[p]}$ of
$\mathbb{M}$ are necessarily, according to Sullivan's theorem, in one--to--one correspondence with
Chevalley $p+1$ cohomology classes $\Gamma^{[p+1]}\left (e\right)$ of $\mathbb{G}\subset \mathbb{M}$.
Indeed, given such a class, which is a polynomial in the $e^I$
generators, we can consistently write the new higher degree Maurer
Cartan equation:
\begin{equation}
  d \, A^{[p]} + \Gamma^{[p+1]}(e) = 0
\label{oppato}
\end{equation}
where $A^{[p]}$ is a new object that cannot be written as a
polynomial in the old objects $e^I$. Considering now the FDA generated
by the inclusion of the available $A^{[p]}$, one can inspect its
Chevalley cohomology: the cochains are the polynomials in the extended set of
forms $\left \{ A,e^I\right \}$ and the boundary operator is defined
by the enlarged set of Maurer Cartan equations. If there are new
cohomology classes $\Gamma^{[p+1]}\left( e,A \right)$, then one can further
extend the FDA by including new $p$--generators $B^{[p]}$ obeying the
Maurer Cartan equation:
\begin{equation}
  \partial \, B^{[p]} + \Gamma^{[p+1]}\left (e,A\right) = 0
\label{oppato2}
\end{equation}
The iterative procedure can now be continued by inspecting the
cohomology classes of type $\Gamma^{[p+1]}\left (e,A,B\right)$ which
lead to new generators $C^{[p]}$ and so on. Sullivan's theorem states that
those constructed in this way are, up to isomorphisms, the most
general minimal FDA.s.
\par
To be precise, this is not the whole story. There is actually one
 generalization that should be taken into account. Instead of
 \textit{absolute Chevalley cohomology} one can rather consider
 \textit{relative Chevalley cohomology}. This means that rather then
being $\mathbb{G}$- singlets, the Chevalley $p$-cochains can be assigned to
some linear representation of the Lie algebra $\mathbb{G}$:
\begin{equation}
  \Omega^{\alpha[p]} = \Omega_{I_1\dots I_p}^\alpha \, e^{I_1} \, \wedge \, \dots
  \wedge \, e^{I_p}
\label{pcochain2}
\end{equation}
where the index $\alpha$ runs in some representation $D$:
\begin{equation}
  D \quad : \quad T_I \, \rightarrow \,
  \left[D\left(T_I\right)\right]^\alpha_{\phantom{\alpha}\beta}
\label{onice}
\end{equation}
 and the boundary operator is now the covariant $\nabla$:
\begin{eqnarray}
  \nabla \, \Omega^{\alpha[p]} & \equiv & \partial \Omega^{\alpha[p]}
  \, +\, e^I \, \wedge \,
  \left[D\left(T_I\right)\right]^\alpha_{\phantom{\alpha}\beta} \,
  \Omega^{\beta[p]}
\label{dcov}
\end{eqnarray}
Since  $\nabla^2=0$, we can repeat all previously explained steps and compute cohomology groups.
Each non trivial cohomology class $\Gamma^{\alpha[p+1]}(e)$ leads to
new $p$--form generators $A^{\alpha[p]}$ which are assigned to the
same $\mathbb{G}$--representation as $\Gamma^{\alpha[p+1]}(e)$. All
successive steps go through in the same way as before and Sullivan's
theorem actually states that all minimal FDA.s are obtained in this
way for suitable choices of the representation $D$, in particular the
singlet.
\section{The super FDA of M theory and its cohomological structure}
\label{superFDA}
Sullivan's theorems have been introduced and proved for Lie algebras
and their corresponding FDA extensions but they hold true, with obvious
modifications, also for superalgebras ${\mathbb{G}}_s$ and for their
FDA extensions. Actually, in view of superstring and supergravity, it
is precisely in the supersymmetric context that FDA.s have found
their most relevant applications. As an illustration
of the general set up let me consider the case of M-theory and of its
FDA, by recalling the results of \cite{fredauria}
and \cite{comments}.
we begin by  writing the complete set of curvatures, plus their Bianchi identities. This
will define the complete FDA:
\begin{equation}
  \mathbb{A}=\mathbb{M}\biguplus\mathbb{C}
\label{completealg}
\end{equation}
The curvatures being the contractible generators $\mathbb{C}$. By setting them to zero we  retrieve,
according to Sullivan's first theorem, the minimal
algebra $\mathbb{M}$. This latter, according instead to Sullivan's second
theorem, has to be explained in terms of cohomology of the
 normal subalgebra $\mathbb{G} \subset \mathbb{M}$, spanned by the
 $1$--forms. In this case  $\mathbb{G}$ is just
  the $D=11$ superalgebra spanned by the following $1$--forms:
\begin{enumerate}
  \item the vielbein $V^a$
  \item the spin connection $\omega^{ab}$
  \item the gravitino $\psi$
\end{enumerate}
The higher degree generators of the minimal FDA $\mathbb{M}$ are:
\begin{enumerate}
  \item the bosonic $3$--form $\mathbf{A^{[3]}}$
  \item the bosonic $6$-form $\mathbf{A^{[6]}}$.
\end{enumerate}
The complete set of curvatures is given below (\cite{fredauria,comments}):
\begin{eqnarray}
T^{a} & = & \mathcal{D}V^a - {\rm i} \ft 12 \, \overline{\psi} \, \wedge \, \Gamma^a \, \psi \nonumber\\
R^{ab} & = & d\omega^{ab} - \omega^{ac} \, \wedge \, \omega^{cb}
\nonumber\\
\rho & = & \mathcal{D}\psi \equiv d \psi - \ft 14 \, \omega^{ab} \, \wedge \, \Gamma_{ab} \, \psi\nonumber\\
\mathbf{F^{[4]}} & = & d\mathbf{A^{[3]}} - \ft 12\, \overline{\psi} \, \wedge \, \Gamma_{ab} \, \psi \,
\wedge \, V^a \wedge V^b \nonumber\\
\mathbf{F^{[7]}} & = & d\mathbf{A^{[6]}} -15 \, \mathbf{F^{[4]}} \, \wedge \,  \mathbf{A^{[3]}} - \ft {15}{2} \,
\, V^{a}\wedge V^{b} \, \wedge \, {\bar \psi}
\wedge \, \Gamma_{ab} \, \psi
\, \wedge \, \mathbf{A^{[3]}} \nonumber\\
\null & \null & - {\rm i}\, \ft {1}{2} \, \overline{\psi} \, \wedge \, \Gamma_{a_1 \dots a_5} \, \psi \,
\wedge \, V^{a_1} \wedge \dots \wedge V^{a_5}
\label{FDAcompleta}
\end{eqnarray}
From their very definition, by taking a further exterior derivative
one obtains the Bianchi identities, which for brevity we do not
explicitly write (see \cite{comments}). The dynamical theory is
defined, according to a general constructive scheme of supersymmetric theories, by the principle
of rheonomy (compare with the tables \ref{flowchart} and \ref{rheotable} ) implemented
into Bianchi identities.
\subsection{Rheonomy}
Indeed there is a unique rheonomic parametrization of the curvatures (\ref{FDAcompleta}) which solves the
Bianchi identities and it  is the following one:
\begin{eqnarray}
T^a & = & 0 \nonumber\\
\mathbf{F^{[4]}} & = & F_{a_1\dots a_4} \, V^{a_1} \, \wedge \dots \wedge \, V^{a_4} \nonumber\\
\mathbf{F^{[7]}} & = & \ft {1}{84} F^{a_1\dots a_4} \, V^{b_1} \, \wedge \dots \wedge \,
V^{b_7} \, \epsilon_{a_1 \dots a_4 b_1 \dots b_7} \nonumber\\
\rho & = & \rho_{a_1a_2} \,V^{a_1} \, \wedge \, V^{a_2} - {\rm i} \ft 12 \,
\left(\Gamma^{a_1a_2 a_3} \psi \, \wedge \, V^{a_4} + \ft 1 8
\Gamma^{a_1\dots a_4 m}\, \psi \, \wedge \, V^m
\right) \, F^{a_1 \dots a_4} \nonumber\\
R^{ab} & = & R^{ab}_{\phantom{ab}cd} \, V^c \, \wedge \, V^d
+ {\rm i} \, \rho_{mn} \, \left( \ft 12 \Gamma^{abmn} - \ft 2 9 \Gamma^{mn[a}\, \delta^{b]c} + 2 \,
\Gamma^{ab[m} \, \delta^{n]c}\right) \, \psi \wedge V^c\nonumber\\
 & &+\overline{\psi} \wedge \, \Gamma^{mn} \, \psi \, F^{mnab} + \ft 1{24} \overline{\psi} \wedge \,
 \Gamma^{abc_1 \dots c_4} \, \psi \, F^{c_1 \dots c_4}
\label{rheoFDA}
\end{eqnarray}
The expressions (\ref{rheoFDA}) satisfy the Bianchi.s provided the space--time components
of the curvatures satisfy the following constraints
\begin{eqnarray}
0 & = & \mathcal{D}_m F^{mc_1 c_2 c_3} \, + \, \ft 1{96} \, \epsilon^{c_1c_2c_3 a_1 a_8} \, F_{a_1 \dots a_4}
\, F_{a_5 \dots a_8}  \nonumber\\
0 & = & \Gamma^{abc} \, \rho_{bc} \nonumber\\
R^{am}_{\phantom{bm}cm} & = & 6 \, F^{ac_1c_2c_3} \,F^{bc_1c_2c_3} -
\, \ft 12 \, \delta^a_b \, F^{c_1 \dots c_4} \,F^{c_1 \dots c_4}
\label{fieldeque}
\end{eqnarray}
which are the  space--time field equations.
\section{The constrained BRST algebra from the FDA}
Applying a general procedure we can obtain the explicit form of
the constrained BRST algebra appropriate to any supergravity.
\par
The first step is the construction of the standard ghost-form extension
of supergravity. This has been codified in the general principle \ref{pippo}
formulated in \cite{Anselmi:1992tj} and recalled in the introduction.
The correct BRST algebra is provided by replacing, in the rheonomic parametrization, of the classical
supergravity curvatures each
differential form with its extended ghost-form counterpart while keeping the curvature components
untouched.
Thus one obtains the rheonomic parametrization of the ghost--extended curvatures, whose formal definition
is identical with that of the classical curvatures upon the replacements
(\ref{orletto1})
\par
The above constructive principle is completely algorithmic and
applies without exception to all supergravity theories. It emphasizes
the fact that the BRST algebra in the positive ghost number sector is,
through, a codified number os steps,
deterministic consequence of the algebraic structure of the FDA,
actually of the original super Lie algebra. Indeed this latter
determines via cohomology its own FDA extension, then the Bianchi
identities determine via rheonomy a unique parametrization of
curvature in superspace and from that we obtain, also uniquely, all
BRST transformations.
\par
The next step consists of equating to zero all bosonic ghosts. This
produces a unique form of a constrained BRST algebra and a unique
form of constraints on the superghosts.
\par
Once the principle has been
clarified we can perform the two steps at once by considering the
purely fermionic ghost-extension and then invoking principle
\ref{pippo}.
\par
Each extended curvature definition $\widehat{\mathbf{R}}^{[p]}_{def}$
and each extended curvature
parametrization $\widehat{\mathbf{R}}^{[p]}_{par}$ decomposes into ghost
sectors according to:
\begin{eqnarray}
\widehat{\mathbf{R}}^{[p]}_{def} & = & {\mathbf{R}}^{[p,0]}_{def} \, + \,
 {\mathbf{R}}^{[p-1,1]}_{def} \, + \, {\mathbf{R}}^{[p-2,2]}_{def} \,\nonumber\\
\widehat{\mathbf{R}}^{[p]}_{par} & = & {\mathbf{R}}^{[p,0]}_{par} \, + \,
 {\mathbf{R}}^{[p-1,1]}_{par} \, + \, {\mathbf{R}}^{[p-2,2]}_{par} \,
\label{curvDefcurvPar}
\end{eqnarray}
where we stop at ghost number $g=2$ since neither in the curvature
definitions nor in the curvature parametrizations there appear higher
than quadratic powers of the $\psi$ forms. Then we have to
impose:
\begin{eqnarray}
{\mathbf{R}}^{[p,0]}_{def} & = & {\mathbf{R}}^{[p,0]}_{par} \nonumber\\
{\mathbf{R}}^{[p-1,1]}_{def} & = & {\mathbf{R}}^{[p-1,1]}_{par}
\nonumber\\
{\mathbf{R}}^{[p-2,2]}_{def} & = & {\mathbf{R}}^{[p-2,2]}_{par}
\label{sectoreque}
\end{eqnarray}
The first of eq.s (\ref{sectoreque}) is simply the rheonomic
parametrization of the classical curvature we started from. The
second equation defines the constrained BRST transformation of all
the physical fields. The last of eq.s (\ref{sectoreque}) defines the
BRST transformation of the ghost fields (the pure spinors) when the
right hand side is non zero ($ {\mathbf{R}}^{[p-2,2]}_{par} \, \ne \,
0$) and the quadratic pure spinor constraints
${\mathbf{R}}^{[p-2,2]}_{def} \, = \, 0$ when the right hand side is
zero ${\mathbf{R}}^{[p-2,2]}_{par}\, = \, 0$.
\par Let us write the result of these straightforward manipulations in the case
of M-theory
\subsection{The constrained BRST algebra of M-theory}
In the case of the minimal FDA of M-theory (disregarding the
$6$-form) the purely fermionic ghost-extension procedure corresponds to setting:
\begin{eqnarray}
V^{a} & \mapsto & V^{\underline{a}} \nonumber\\
\mathbf{A}^{[3]} & \mapsto & \mathbf{A}^{[3]} \nonumber\\
\psi  & \mapsto & \psi  \, + \, \lambda
\label{constghostextM}
\end{eqnarray}
where $\lambda$ is the commuting superghosts.
Next it
is convenient to introduce a Lorentz covariant formalism by splitting
the ghost extended Lorentz covariant derivative in the following way:
\begin{eqnarray}
\widehat{D} & = & \widehat{d} \, + \,\widehat{ \omega} ^{ab} \, J_{ab} \nonumber\\
\null & = & d \, + \, s \, + \, \omega^{ab}\, J_{ab} \,+ \, \epsilon^{ab}\, J_{ab} \nonumber\\
\null & = & \mathcal{D} \, + \,  \mathcal{S}\nonumber\\
\mbox{where} & \null & \null \nonumber\\
\mathcal{D} & = & d \, +\, \omega^{ab}\, J_{ab} \quad \mbox{Lorentz covariant external
derivative}\nonumber\\
\mathcal{S} & = & s \, + \,\epsilon^{ab}\, J_{ab}\quad \mbox{Lorentz covariant
BRST variation}
\label{Lorenzocovari}
\end{eqnarray}
and where $J_{ab}$ denotes the standard generators of the $\mathrm{SO(1,10)}$ Lie
algebra. In the above formulae $\epsilon^{ab}$ are the Lorentz-ghosts
which are field dependent on the superghosts in the usual way as the
spin connection is field dependent on the gravitinos upon solving the
zero torsion equation.
\par
With these notations, from the second of equations (\ref{sectoreque})
we obtain the BRST-transformations of the physical fields:
\begin{eqnarray}
\mathcal{S}  V^a &   = & {\rm i} \, \overline{\Psi} \, \Gamma^a \, \lambda \, \label{SdiVa}
\nonumber \\
 \mathcal{S} \,\Psi &  = & - \, \, \mathcal{D}\, \lambda \, - \, {\rm i} \ft 12 \,
\left(\Gamma^{a_1a_2 a_3} \lambda \,  V^{a_4} + \ft 1 8
\Gamma^{a_1\dots a_4 m}\, \lambda \,  V^m
\right) \, F^{a_1 \dots a_4}\nonumber\\
\null & \equiv & - \nabla \, \lambda \label{Sdipsi}
\nonumber
\\
 \mathcal{S} \, \mathbf{A^{[3]}}  & = &   \, \overline{\Psi} \, \wedge \,
 \Gamma_{ab} \, \lambda \,
\wedge \, V^a \wedge V^b
\label{phystransfor2}
\end{eqnarray}
while from the third of eq.s (\ref{sectoreque}) we obtain the BRST
variation of the superghost and the appropriate pure spinor
constraints, namely:
\begin{eqnarray}
\mathcal{S} \lambda &  = & \,0 \nonumber\\
 0 & = & \overline{\lambda} \, \Gamma^a \, \lambda\nonumber\\
0 & = & \overline{\lambda}  \, \Gamma^{mn} \, \lambda
\, V_m \, \wedge \, V_n
\label{hatalgebra}
\end{eqnarray}
In addition from the rheonomic parametrization of the Lorentz curvatures we also
learn the closure relations satisfied by the commutators and
anticommutators of the operators $\mathcal{D}$ and $\mathcal{S}$.
They are as follows:
\begin{equation}
  \begin{array}{rcl}
    \mathcal{S}^2 & = & \begin{array}{l}
     \big [  \overline{\lambda}  \, \Gamma^{mn} \, \lambda \, F^{mnab} + \ft 1{24} \overline{\lambda}  \,
 \Gamma^{abc_1 \dots c_4} \, \lambda \, F^{c_1 \dots c_4} \big ] \, J_{ab}\\
    \end{array} \\
\mathcal{D}^2 & = & \begin{array}{l}
    \null\\
    \null\\
     \big [ R^{ab}_{\phantom{ab}mn} \, V^m \, \wedge V^{n} \, \\
      + \,{\rm i} \, {\bar \rho}_{mn} \, \left( \ft 12 \Gamma^{abmn} - \ft 2 9 \Gamma^{mn[a}\, \delta^{b]c} + 2 \,
\Gamma^{ab[m} \, \delta^{n]c}\right) \,  \Psi \,  \wedge \, V^c \\
      \, + \, \overline{\Psi}  \, \wedge \, \Gamma^{mn} \, \Psi \, F^{mnab} + \ft 1{24} \overline{\Psi}
      \, \wedge \,
 \Gamma^{abc_1 \dots c_4} \, \Psi \, F^{c_1 \dots c_4} \big ] \, J_{ab}\\
    \end{array} \\
    \mathcal{S}\, \mathcal{D} \, + \,  \mathcal{D}\, \mathcal{S} \, & = & \begin{array}{l}\null\\
     \big [
\,{\rm i} \, {\bar \rho}_{mn} \, \left( \ft 12 \Gamma^{abmn} - \ft 2 9 \Gamma^{mn[a}\, \delta^{b]c} + 2 \,
\Gamma^{ab[m} \, \delta^{n]c}\right) \,  \lambda   \wedge \, V^c \\
      + \, 2 \, \overline{\lambda}  \, \Gamma^{mn} \, \Psi \, F^{mnab} + \ft 1{12} \overline{\lambda}
      \,
 \Gamma^{abc_1 \dots c_4} \, \Psi \, F^{c_1 \dots c_4} \big ] \, J_{ab} \
  \end{array}\\
  \end{array}
\label{algebrona}
\end{equation}
\subsection{Discussion on the constraints}
We have seen that the pure constraints coming from the FDA algebra are
\begin{eqnarray}
0  =  \overline{\lambda} \, \Gamma^m \, \lambda\,, ~~~~~~~~
0  =  \overline{\lambda}  \, \Gamma^{mn} \, \lambda
\, V_m \, \wedge \, V_n\,.
\label{hatalgebra2}
\end{eqnarray}
We can decompose them into irreducible representations of $\mathrm{SO(1,9)}$ and in that
case we have
\begin{eqnarray}
0  =  \overline{\lambda} \, \Gamma^I \, \lambda\,, &&
0  =  \overline{\lambda} \, \Gamma^{11} \, \lambda\,,\nonumber\\
0 =  \overline{\lambda}  \, \Gamma^{IJ} \, \lambda \, V_I \, \wedge \, V_J\,, &&
0  =  \overline{\lambda}  \, \Gamma^{I 11} \, \lambda \, V_I
\label{hatalgebra3}
\end{eqnarray}
where the indices $I,J$ run from 0 to 9. In this way we can analyze the difference between these
constraints and the pure spinor constraints in the case of type IIA superstrings
\cite{Berkovits:2001ue}. In work \cite{Berkovits:2002uc} only the first set of constraints have been analyzed
 yielding  23 independent d.o.f. for the pure spinor field $\lambda$. However, the second
type of constraints becomes necessary  to ensure the BRST invariance of the M2 action.
In \cite{Berkovits:2002uc} a stronger form of the constraint appeared, but in the present derivation we find only
 (\ref{hatalgebra2}).
\par
We would like to study the number of independent d.o.f. from (\ref{hatalgebra2}). For that we use the Fierz identity
for the gamma matrices in 11d and we get the relations (decomposed in the $\mathrm{SO(1,9)}$ representations)
\begin{eqnarray}\label{deco10}
 \overline{\lambda} \, \Gamma^{IJ} \, \lambda\,  \overline{\lambda} \, \Gamma_J \, \lambda +
 \overline{\lambda} \, \Gamma^{I 11} \, \lambda\,  \overline{\lambda} \, \Gamma_{11}\, \lambda=0 \,,
 ~~~~~~~
\overline{\lambda} \, \Gamma^{11 J} \, \lambda\,  \overline{\lambda} \, \Gamma_J \, \lambda =0
 \end{eqnarray}
Using the first constraint $\overline{\lambda} \, \Gamma_J \, \lambda =0$, we reduced them to the equation
\begin{eqnarray}
\overline{\lambda} \, \Gamma^{I 11} \, \lambda\,  \overline{\lambda} \, \Gamma_{11}\, \lambda=0 \,.
\end{eqnarray}
We notice that by contracting the free index $I$ with the 10d vielbein $V_I$, we do not get any condition
on $\overline{\lambda} \, \Gamma_{11}\, \lambda$.
However, if we assume that
$\overline{\lambda} \, \Gamma^{11 J} \, \lambda\, \neq 0$ for $J$ orthogonal to $V_J$, we get that
$\overline{\lambda} \, \Gamma_{11}\, \lambda =0$ as a consequence.
\par
Using a decomposition of the spinors $\lambda, \overline\lambda$ in terms of an adapted basis it can be shown that
equations (\ref{deco10}) implies that there are 22 independent d.o.f and this coincides with the counting
of the pure spinor constraints for type IIA superstrings \cite{Berkovits:2001ue}. It is impressive that even if
the structure of the constraints are different, the number of independent d.o.f. is the same. Therefore, it seems
that by exploiting the complete set of constraints for the supermembrane (namely the pure spinor constraints
(\ref{hatalgebra2})) one finds he agreement with the pure spinor constraints for the superstrings. It will be subject
of another paper the complete discussion on the pure spinor constraints for type IIA/B in presence of RR
fields from the FDA algebras \cite{pie_ant3}.
\section{Conclusions}
In this communication the main goal has been that of illustrating the
intimate relation between the pure spinor BRST algebra and the
structure of the FDA underlying supergravity. This sheds new light
on the quantization of superstrings $\mbox{\'a}$ la pure spinors. This
latter appears the only viable candidate to include the coupling of
the string to Ramond-Ramond background fields.

\end{document}